\documentclass[preprintnumbers,amsmath,amssymbm,prd]{revtex4}
\usepackage{epsfig}
\usepackage{graphicx}
\usepackage{amssymb}

\begin{document}
\title{The superradiant instability regime of the spinning Kerr black hole}
%\title{A bound on the superradiant instability regime of the spinning Kerr black hole}
%\title{The rotating Kerr black-hole bomb: An upper bound on the mass of the explosive scalar field}
\author{Shahar Hod}
\affiliation{The Ruppin Academic Center, Emeq Hefer 40250, Israel}
\affiliation{ } \affiliation{The Hadassah Institute, Jerusalem
91010, Israel}
\date{\today}

\begin{abstract}
Spinning Kerr black holes are known to be superradiantly unstable to
massive scalar perturbations. We here prove that the instability
regime of the composed Kerr-black-hole-massive-scalar-field system
is bounded from above by the dimensionless inequality $M\mu<m\cdot
\sqrt{{{2(1+\gamma)(1-\sqrt{1-\gamma^2})-\gamma^2}\over{4\gamma^2}}}$,
where $\{\mu,m\}$ are respectively the proper mass and azimuthal
harmonic index of the scalar field and $\gamma\equiv r_-/r_+$ is the
dimensionless ratio between the horizon radii of the black hole. It
is further shown that this {\it analytically} derived upper bound on
the superradiant instability regime of the spinning Kerr black hole
agrees with recent {\it numerical} computations of the instability
resonance spectrum.
\end{abstract}
\bigskip
\maketitle

\section {Introduction}

The intriguing physical mechanism of superradiance
\cite{Zel,PressTeu2,Viln} allows an incident bosonic wave field to
extract rotational energy from a spinning Kerr black hole. In
particular, a scalar field mode of azimuthal harmonic index $m$ can
be amplified (that is, can gain energy) as it scatters off a Kerr
black hole if its proper frequency $\omega_{\text{field}}$ lies in
the bounded regime \cite{Zel,PressTeu2,Viln,Noteun}
\begin{equation}\label{Eq1}
0<\omega_{\text{field}}<m\Omega_{\text{H}}\  ,
\end{equation}
where \cite{Chan,Kerr,Notesmp}
\begin{equation}\label{Eq2}
\Omega_{\text{H}}={{a}\over{r^2_++a^2}}\
\end{equation}
is the angular velocity of the spinning Kerr black hole (Here $a$
and $r_+$ are respectively the angular momentum per unit mass and
the outer horizon-radius of the Kerr black hole).

What is even more remarkable is the fact that the rate of energy
extraction from the spinning Kerr black hole in the superradiant
regime (\ref{Eq1}) can grow exponentially in time if the scattered
scalar wave field, which is used to extract the black-hole
rotational energy, is prevented from radiating its energy to
infinity. Interestingly, the Klein-Gordon wave equation for a scalar
field of mass $\mu$ \cite{Notemas,Hodrc,HerR,Notedim} in the Kerr
black-hole spacetime is governed by an effective binding potential
[see Eqs. (\ref{Eq18}) and (\ref{Eq19}) below] which provides a
natural confinement mechanism that prevents low frequency field
modes in the regime
\begin{equation}\label{Eq3}
0<\omega_{\text{field}}<\mu\
\end{equation}
from escaping to infinity. Scalar field modes which respect the
inequalities (\ref{Eq1}) and (\ref{Eq3}) in the rotating Kerr
black-hole spacetime can grow exponentially over time
\cite{Notemas}, thus leading to the formation of a composed
Kerr-black-hole-massive-scalar-field bomb \cite{Notemir,CarDias}.

The boundary between stable ($\omega>m\Omega_{\text{H}}$) and
unstable ($\omega<m\Omega_{\text{H}}$) composed
Kerr-black-hole-massive-scalar-field systems is marked by the
presence of {\it stationary} field configurations whose orbital
frequencies are in resonance with the angular velocity
$\Omega_{\text{H}}$ of the spinning black hole \cite{Hodrc,HerR}.
Specifically, for a given value of the field azimuthal harmonic
index $m$, these marginally-stable (stationary) bound-state field
configurations are characterized by the resonance relation
\cite{Hodrc,HerR}
\begin{equation}\label{Eq4}
\omega_{\text{field}}=\omega_{\text{c}}\equiv m\Omega_{\text{H}}\  ,
\end{equation}
where $\omega_{\text{c}}$ is the critical (threshold) frequency for
superradiant scattering in the Kerr black-hole spacetime.

It was previously proved \cite{Hodbound,Notebf} that, for a scalar
field of proper mass $\mu$ interacting with a spinning Kerr black
hole of angular velocity $\Omega_{\text{H}}$, the inequality
\begin{equation}\label{Eq5}
\mu<\sqrt{2}\cdot m\Omega_{\text{H}}\
\end{equation}
provides an upper bound on the domain of existence of stationary
Kerr-black-hole-massive-scalar-field configurations. Since these
stationary (marginally-stable) field configurations mark the
boundary between stable and unstable Kerr-massive-scalar-field
systems, the relation (\ref{Eq5}) also provides an upper bound on
the superradiant instability regime of the composed
Kerr-black-hole-massive-scalar-field system.

The main goal of the present paper is to derive a {\it stronger}
upper bound on the superradiant instability regime of the spinning
Kerr black-hole spacetime \cite{Notebst}. In particular, below we
shall show that the binding potential well, which is required in
order to support the stationary (marginally-stable) scalar field
configurations (\ref{Eq4}) in the rotating Kerr black-hole
spacetime, exists only in a restricted regime
$\mu/m\Omega_{\text{H}}<{\cal F}(\gamma)$ \cite{Notega} of the
black-hole-field physical parameters. Since this inequality sets an
upper bound on the domain of existence of these marginally-stable
(stationary \cite{Notebst}) field configurations in the rotating
Kerr black-hole spacetime, it also sets an upper bound on the
superradiant instability regime of the composed
Kerr-black-hole-massive-scalar-field system.

\section{Description of the system}

We shall study the dynamics of a massive scalar field $\Psi$ which
is linearly coupled to a spinning Kerr black hole. The black-hole
spacetime is described by the line element \cite{Chan,Kerr}
\begin{eqnarray}\label{Eq6}
ds^2=-{{\Delta}\over{\rho^2}}(dt-a\sin^2\theta
d\phi)^2+{{\rho^2}\over{\Delta}}dr^2+\rho^2
d\theta^2+{{\sin^2\theta}\over{\rho^2}}\big[a
dt-(r^2+a^2)d\phi\big]^2\  ,
\end{eqnarray}
where $(t,r,\theta,\phi)$ are the Boyer-Lindquist coordinates,
$\{M,a\}$ are the mass and angular momentum per unit mass of the
black hole, and
\begin{equation}\label{Eq7}
\Delta\equiv r^2-2Mr+a^2\ \ \ ; \ \ \ \rho^2\equiv
r^2+a^2\cos^2\theta\  .
\end{equation}
The zeros of $\Delta$,
\begin{equation}\label{Eq8}
r_{\pm}=M\pm\sqrt{M^2-a^2}\  ,
\end{equation}
are the (outer and inner) horizon radii of the spinning black hole.

The dynamics of a linearized scalar field $\Psi$ of proper mass
$\mu$ in the black-hole spacetime is governed by the Klein-Gordon
wave equation
\begin{equation}\label{Eq9}
(\nabla^{\nu}\nabla_{\nu}-\mu^2)\Psi=0\  .
\end{equation}
One can decompose the eigenfunction $\Psi$ of the massive scalar
field in the form \cite{Noteanz}
\begin{equation}\label{Eq10}
\Psi(t,r;\omega,\theta,\phi)=\sum_{l,m}e^{im\phi}{S_{lm}}(\theta;m,a\sqrt{\mu^2-\omega^2})
{R_{lm}}(r;M,a,\mu,\omega)e^{-i\omega t}\  .
\end{equation}
Substituting (\ref{Eq10}) into the Klein-Gordon wave equation
(\ref{Eq9}), one finds that the angular function $S_{lm}$ satisfies
the angular equation \cite{Stro,Heun,Fiz1,Teuk,Abram,Hodasy}
\begin{eqnarray}\label{Eq11}
{1\over {\sin\theta}}{{d}\over{d\theta}}\Big(\sin\theta {{d
S_{lm}}\over{d\theta}}\Big)
+\Big[K_{lm}+a^2(\mu^2-\omega^2)\sin^2\theta-{{m^2}\over{\sin^2\theta}}\Big]S_{lm}=0\
.
\end{eqnarray}
Demanding the angular functions to be regular at the two poles
$\theta=0$ and $\theta=\pi$, one finds that the differential
equation (\ref{Eq11}) is characterized by a discrete set
$\{K_{lm}\}$ of angular eigenvalues (see \cite{Barma,Yang,Hodpp} and
references therein). Below we shall use the fact that the
characteristic eigenvalues of the angular equation (\ref{Eq11}) are
bounded from below by the relation \cite{Barma,Notesi}
\begin{equation}\label{Eq12}
K_{lm}\geq m^2-a^2(\mu^2-\omega^2)\  .
\end{equation}

The radial function $R_{lm}$ satisfies the radial equation
\cite{Teuk,Stro}
\begin{equation}\label{Eq13}
\Delta{{d} \over{dr}}\Big(\Delta{{d R_{lm}
}\over{dr}}\Big)+\Big\{[\omega(r^2+a^2)-ma]^2
+\Delta[2ma\omega-\mu^2(r^2+a^2)-K_{lm}]\Big\}R_{lm}=0\ .
\end{equation}
It is worth noting that the angular eigenvalues $\{K_{lm}\}$ couple
equation (\ref{Eq13}) for the radial eigenfunctions to equation
(\ref{Eq11}) for the angular eigenfunctions \cite{Notenam}. The
radial equation (\ref{Eq13}) should be supplemented by the physical
boundary condition of purely ingoing waves (as measured by a
comoving observer) at the horizon of the black hole
\cite{Notemas,Hodrc,HerR}:
\begin{equation}\label{Eq14}
R_{lm} \sim e^{-i(\omega-m\Omega_{\text{H}})y}\ \ \ \text{ for }\ \
\ r\rightarrow r_+\ \ (y\rightarrow -\infty)\ ,
\end{equation}
where the radial coordinate $y$ is determined by the relation
$dy=(r^2/\Delta)dr$ [see Eq. (\ref{Eq17}) below]. In addition, the
asymptotic (large-$r$) behavior \cite{Notemas,Hodrc,HerR}
\begin{equation}\label{Eq15}
R_{lm} \sim {1\over r}e^{-\sqrt{\mu^2-\omega^2}y}\ \ \ \text{ for }\
\ \ r\rightarrow\infty\ \ (y\rightarrow \infty)\
\end{equation}
of the radial eigenfunction, together with the characteristic
inequality (\ref{Eq3}), guarantee that the external bound-state
configurations of the massive scalar fields are characterized by
spatially decaying (bounded) radial eigenfunctions at asymptotic
infinity.

\section{The effective binding potential of the composed Kerr-black-hole-massive-scalar-field system}

Our main goal is to obtain an upper bound on the domain of existence
of the stationary (marginally-stable)
Kerr-black-hole-massive-scalar-field configurations \cite{Notebst}.
To this end, it proves useful to transform the radial equation
(\ref{Eq13}) into a Schr\"odinger-like wave equation. Substituting
\begin{equation}\label{Eq16}
\psi=rR\
\end{equation}
and \cite{Notemap}
\begin{equation}\label{Eq17}
dy={{r^2}\over{\Delta}}dr\
\end{equation}
into the radial equation (\ref{Eq13}), one obtains the
Schr\"odinger-like wave equation
\begin{equation}\label{Eq18}
{{d^2\psi}\over{dy^2}}-V(y)\psi=0\  ,
\end{equation}
where the effective potential which governs the radial equation
(\ref{Eq18}) is given by
\begin{equation}\label{Eq19}
V=V(r;\omega,M,a,\mu,l,m)={{2\Delta}\over{r^6}}(Mr-a^2)+{{\Delta}\over{r^4}}
[K_{lm}-2ma\omega+\mu^2(r^2+a^2)]-{{1}\over{r^4}}[\omega(r^2+a^2)-ma]^2\
.
\end{equation}
Note that this radial potential is characterized by the asymptotic
properties [see Eqs. (\ref{Eq2}), (\ref{Eq4}), and (\ref{Eq19})]
\begin{equation}\label{Eq20}
V(r=r_+;\omega=\omega_{\text{c}},M,a,\mu,l,m)=0\
\end{equation}
and
\begin{equation}\label{Eq21}
V(r\to \infty;\omega=\omega_{\text{c}},M,a,\mu,l,m)\to
\mu^2-\omega^2_{\text{c}}>0
\end{equation}
at the black-hole horizon and at spatial infinity, respectively.

In the next section we shall analyze the spatial properties of the
effective radial potential (\ref{Eq19}) for the stationary
\cite{Notebst} bound-state configurations of the massive scalar
fields in the rotating Kerr black-hole spacetime [these marginally
stable field configurations are characterized by the critical
(threshold) superradiant frequency (\ref{Eq4})]. In particular, we
shall show that the requirement that the effective radial potential
(\ref{Eq19}) has the form of a binding potential well outside the
black-hole horizon sets an upper bound on the mass of the explosive
scalar field in the superradiant regime (\ref{Eq1}).

\section{An upper bound on the mass of the explosive scalar field}

In the present section we shall study the spatial behavior of the
effective radial potential (\ref{Eq19}) which governs the
interaction of the stationary (marginally-stable) massive scalar
configurations with the rotating Kerr black hole. Substituting the
characteristic resonant frequency (\ref{Eq4}) of the stationary
scalar fields into the expression (\ref{Eq19}) for the effective
radial potential, and using the inequality (\ref{Eq12}) for the
eigenvalues of the angular equation (\ref{Eq11}), one obtains the
inequality
\begin{equation}\label{Eq22}
V(r)\geq m^2\cdot {{r-r_+}\over{r^3(r^2_++a^2)^2}}\big[\beta
a^2r^2-a^2(2M+\beta
r_-)r+2Mr^3_+\big]+{{2\Delta}\over{r^6}}(Mr-a^2)\
\end{equation}
for the effective radial potential which characterizes the
stationary Kerr-black-hole-massive-scalar-field configurations,
where the dimensionless parameter $\beta>0$ \cite{Notebet} is
defined by the relation
\begin{equation}\label{Eq23}
\mu^2=(1+\beta)\cdot\omega^2_{\text{c}}\  .
\end{equation}
Furthermore, substituting the inequality $Mr-a^2\geq0$
\cite{Noteineq} into (\ref{Eq22}), one finds the lower bound
\begin{equation}\label{Eq24}
V(r)\geq m^2\cdot {{r-r_+}\over{r^3(r^2_++a^2)^2}}\big[\beta
a^2r^2-a^2(2M+\beta r_-)r+2Mr^3_+\big]
\end{equation}
on the effective radial potential.

A necessary condition for the existence of the stationary
(marginally-stable) bound-state scalar configurations in the
rotating Kerr black-hole spacetime is provided by the requirement
that the effective radial potential (\ref{Eq19}), which
characterizes the composed Kerr-black-hole-massive-scalar-field
system, has the form of a {\it binding} potential well. In
particular, taking cognizance of the property (\ref{Eq20}) which
characterizes the effective radial potential of the composed
black-hole-field system, one concludes that the inequality
\begin{equation}\label{Eq25}
V(r)\leq0\ \ \ \ \text{for}\ \ \ \ r\in
[r^-_{\text{b}},r^+_{\text{b}}]\
\end{equation}
provides a necessary condition for the existence of stationary
bound-state scalar configurations in the Kerr black-hole spacetime.
The inequality (\ref{Eq25}) reflects the fact that, in order to be
able to support stationary bound-state scalar configurations outside
the black-hole horizon, the effective radial potential (\ref{Eq19})
of the composed black-hole-field system must have the form of a
binding potential well in some interval $r_+\leq r^-_{\text{b}}\leq
r\leq r^+_{\text{b}}$ outside the black-hole horizon.

Taking cognizance of Eqs. (\ref{Eq24}) and (\ref{Eq25}), one finds
the characteristic inequality
\begin{equation}\label{Eq26}
\beta a^2r^2-a^2(2M+\beta r_-)r+2Mr^3_+\leq0\
\end{equation}
in the interval $r\in [r^-_{\text{b}},r^+_{\text{b}}]$ which
characterizes the binding potential well outside the black-hole
horizon. The zeros of the quadratic function on the l.h.s of
(\ref{Eq26}) are given by
\begin{equation}\label{Eq27}
r^{\pm}_{\text{b}}={{a(2M+\beta r_-)\pm\sqrt{a^2(2M+\beta
r_-)^2-8\beta Mr^3_+}}\over{2\beta a}}\ .
\end{equation}
The requirements $r^{\pm}_{\text{b}}\in \mathbb{R}$ with $r_+\leq
r^-_{\text{b}}\leq r^+_{\text{b}}$ yield the relation
%\begin{equation}\label{Eq32}
%a^2(2M+\beta r_-)^2-8\beta Mr^3_+\geq 0
%\end{equation}
\begin{equation}\label{Eq28}
a^2r^2_-\cdot\beta^2+4M(a^2r_--2r^3_+)\cdot\beta+4M^2a^2>0
\end{equation}
as a necessary condition for the validity of the inequality
(\ref{Eq26}). From (\ref{Eq28}) one finds the upper bound
\begin{equation}\label{Eq29}
\beta<
{{2M\big[2r^2_+-r^2_--2r_+({r^2_+-r^2_-})^{1/2}\big]}\over{r^3_-}}
\end{equation}
on the dimensionless quantity $\beta$.

Finally, taking cognizance of Eqs. (\ref{Eq4}), (\ref{Eq23}), and
(\ref{Eq29}), and defining the dimensionless ratio
\begin{equation}\label{Eq30}
\gamma\equiv {{r_-}\over{r_+}}\
\end{equation}
between the horizon radii of the spinning Kerr black hole, one finds
the upper bound
\begin{equation}\label{Eq31}
\mu< {\cal F}(\gamma)\cdot m\Omega_{\text{H}}
\end{equation}
on the scalar mass of the stationary (marginally-stable) bound-state
field configurations, where the dimensionless function ${\cal
F}={\cal F}(\gamma)$ is given by
\begin{equation}\label{Eq32}
{\cal
F}(\gamma)=\sqrt{{{2(1+\gamma)(1-\sqrt{1-\gamma^2})-\gamma^2}\over{\gamma^3}}}\
.
\end{equation}

It is worth emphasizing again that the stationary bound-state field
configurations (\ref{Eq4}) mark the onset of the superradiant
instabilities in the composed Kerr-black-hole-massive-scalar-field
system. Thus, the analytically derived upper bound (\ref{Eq31}) on
the domain of existence of these stationary (marginally-stable)
black-hole-field configurations also provides an upper bound on the
superradiant instability regime of the composed
Kerr-black-hole-massive-scalar-field system \cite{Notebt}.

\section{Numerical confirmation}

It is of physical interest to test the validity of the analytically
derived upper bound (\ref{Eq31}) on the superradiant instability
regime of the composed Kerr-black-hole-massive-scalar-field system.
As emphasized earlier, the boundary between stable and unstable
composed black-hole-field systems is marked by the stationary
(marginally-stable) black-hole-field configurations studied in
\cite{Hodrc,HerR}. In particular, the scalar field masses
$\mu=\mu(m,\Omega_{\text{H}})$ which correspond to these stationary
(marginally-stable) composed Kerr-black-hole-massive-scalar-field
configurations were computed numerically in \cite{HerR}.

In Table \ref{Table1} we present the dimensionless ratio
${{\mu_{\text{numerical}}}/{\mu_{\text{bound}}}}$, where
$\mu_{\text{numerical}}$ is the numerically computed \cite{HerR}
field masses which mark the onset of the superradiant instabilities
in the composed Kerr-black-hole-massive-scalar-field system, and
$\mu_{\text{bound}}$ is the analytically derived upper bound
(\ref{Eq31}) on the superradiant instability regime of the composed
black-hole-field system. One finds from Table \ref{Table1} that the
superradiant instability regime
%stationary (marginally-stable) configurations
of the composed Kerr-black-hole-massive-scalar-field system is
characterized by the relation
${{\mu_{\text{numerical}}}/{\mu_{\text{bound}}}}<1$, in agreement
with the analytically derived upper bound (\ref{Eq31}).

\begin{table}[htbp]
\centering
\begin{tabular}{|c|c|c|c|}
\hline \ \ $s\equiv {{a}/{M}}$\ \ & \ ${\cal F}(s)$\ \ & \ \
${{\mu(l=m=1)}\over{\mu_{\text{bound}}}}$\ \ & \ \ ${{\mu(l=m=10)}\over{\mu_{\text{bound}}}}$\ \ \\
\hline
\ \ 0.1\ \ & \ \ 1.00031\ \ \ & \ \ 0.99977\ \ \ & \ \ 0.99940\ \ \ \\
\ \ 0.2\ \ & \ \ 1.00129\ \ \ & \ \ 0.99903\ \ \ & \ \ 0.99967\ \ \ \\
\ \ 0.3\ \ & \ \ 1.00301\ \ \ & \ \ 0.99774\ \ \ & \ \ 0.99948\ \ \ \\
\ \ 0.4\ \ & \ \ 1.00567\ \ \ & \ \ 0.99573\ \ \ & \ \ 0.99901\ \ \ \\
\ \ 0.5\ \ & \ \ 1.00960\ \ \ & \ \ 0.99276\ \ \ & \ \ 0.99776\ \ \ \\
\ \ 0.6\ \ & \ \ 1.01541\ \ \ & \ \ 0.98835\ \ \ & \ \ 0.99715\ \ \ \\
\ \ 0.7\ \ & \ \ 1.02437\ \ \ & \ \ 0.98163\ \ \ & \ \ 0.99558\ \ \ \\
\ \ 0.8\ \ & \ \ 1.03955\ \ \ & \ \ 0.96995\ \ \ & \ \ 0.99276\ \ \ \\
\ \ 0.9\ \ & \ \ 1.07168\ \ \ & \ \ 0.94694\ \ \ & \ \ 0.98676\ \ \ \\
\ \ 0.95\ \ & \ \ 1.11039\ \ \ & \ \ 0.91878\ \ \ & \ \ 0.97942\ \ \ \\
\ \ 0.99\ \ & \ \ 1.21646\ \ \ & \ \ 0.84910\ \ \ & \ \ 0.96165\ \ \ \\
\ \ 0.999\ \ & \ \ 1.37370\ \ \ & \ \ 0.76084\ \ \ & \ \ 0.94280\ \ \ \\
\hline
\end{tabular}
\caption{The superradiant instability regime of the composed
Kerr-black-hole-massive-scalar-field system (the rotating black-hole
bomb). We present the dimensionless ratio
${{\mu_{\text{numerical}}}/{\mu_{\text{bound}}}}$, where
$\mu_{\text{numerical}}$ is the numerically computed \cite{HerR}
field masses which mark the onset of the superradiant instabilities
in the composed Kerr-black-hole-massive-scalar-field system, and
$\mu_{\text{bound}}$ is the analytically derived upper bound on the
superradiant instability regime of the composed system as given by
Eq. (\ref{Eq31}). The data presented are for the cases $l=m=1$ and
$l=m=10$. One finds that the superradiant instability regime of the
composed Kerr-black-hole-massive-scalar-field system is
characterized by the relation
${{\mu_{\text{numerical}}}/{\mu_{\text{bound}}}}<1$, in agreement
with the analytically derived upper bound (\ref{Eq31}).}
\label{Table1}
\end{table}

\section{Summary}

In this paper we have explored the superradiant instability regime
of the spinning Kerr black hole to massive scalar perturbations. In
particular, we have shown that the binding potential well, which is
required in order to support the stationary \cite{Notebst}
bound-state scalar field configurations (\ref{Eq4}) in the rotating
black-hole spacetime, exists only in the restricted regime [see Eqs.
(\ref{Eq31}) and (\ref{Eq32})]
\begin{equation}\label{Eq33}
%M\mu<m\cdot
%{{\sqrt{2(1+\gamma)(1-\sqrt{1-\gamma^2})-\gamma^2}}\over{2\gamma}}\
%.
M\mu<m\cdot
\sqrt{{{2(1+\gamma)(1-\sqrt{1-\gamma^2})-\gamma^2}\over{4\gamma^2}}}\
\ \ \ \ ; \ \ \ \ \gamma\equiv r_-/r_+\  .
\end{equation}

The dimensionless inequality (\ref{Eq33}) sets an upper bound on the
domain of existence of the stationary bound-state scalar field
configurations (\ref{Eq4}) in the rotating Kerr black-hole
spacetime. Since these marginally-stable (stationary) bound-state
scalar field configurations \cite{Notesf} mark the boundary between
stable ($\omega>m\Omega_{\text{H}}$) and unstable
($\omega<m\Omega_{\text{H}}$) field configurations in the Kerr
black-hole spacetime, the analytically derived inequality
(\ref{Eq33}) also provides an upper bound on the superradiant
instability regime \cite{Notenecs} of the composed
Kerr-black-hole-massive-scalar-field system \cite{Notefs}.

\newpage
\bigskip
\noindent
{\bf ACKNOWLEDGMENTS}
\bigskip

This research is supported by the Carmel Science Foundation. I would
like to thank C. A. R. Herdeiro and E. Radu for sharing with me
their numerical data. I thank Yael Oren, Arbel M. Ongo, Ayelet B.
Lata, and Alona B. Tea for stimulating discussions.

%\newpage


\begin{thebibliography}{99}
\bibitem{Noteun} We use natural units in which $G=c=\hbar=1$.
\bibitem{Notesmp} We shall assume $a>0$ and $m>0$ without loss of
generality.
\bibitem{Notedim} Note that the characteristic mass parameter $\mu$ of the scalar field stands
for $\mu/\hbar$. Hence, this physical parameter has the dimensions
of $($length$)^{-1}$.
\bibitem{Notemir} As shown in \cite{PressTeu2,CarDias}, an artificial confinement mechanism
can be provided by a reflecting mirror which surrounds the black
hole and prevents the amplified scalar field modes from radiating
their energy to infinity. This composed setup is known as the
Kerr-black-hole-scalar-field-mirror bomb.
\bibitem{Notebst} It is worth emphasizing again that the onset of
the superradiant instabilities in the composed
Kerr-black-hole-massive-scalar-field system is marked by the
presence of the stationary (marginally-stable) bound-state field
configurations (\ref{Eq4}). Thus, an upper bound on the domain of
existence of these stationary black-hole-field configurations [see
Eq. (\ref{Eq31}) below] would also provide an upper bound on the
superradiant instability regime of the composed
Kerr-black-hole-massive-scalar-field system.
\bibitem{Notega} The dimensionless scaling function
${\cal F}(\gamma)$, to be determined below, depends solely on the
dimensionless ratio $\gamma\equiv {{r_-}/{r_+}}$ between the horizon
radii of the spinning Kerr black hole [see Eq. (\ref{Eq32}) below].
\bibitem{Noteanz} The integer parameters $m$ and $l\geq|m|$ are
the (azimuthal and spheroidal) harmonic indices of the scalar field
modes [see Eq. (\ref{Eq11}) below].
\bibitem{Notesi} It is worth noting that the lower bound (\ref{Eq12}) on the angular eigenvalues can be saturated in
the eikonal $l=m\gg1$ limit, in which case one finds the simple
relation $K_{mm}=m^2[1+O(m^{-1})]-a^2(\mu^2-\omega^2)$ \cite{Hodpp}.
\bibitem{Notenam} Note, in particular, that the eigenvalues
$\{K_{lm}\}$ which characterize the angular equation (\ref{Eq11})
also appear in the effective potential of the radial Klein-Gordon
equation (\ref{Eq13}).
\bibitem{Notemap} Note that the coordinate $r\in [0,\infty]$ is mapped
into $y\in [-\infty,+\infty]$ by the radial transformation
(\ref{Eq17}).
\bibitem{Notebet} Note that the inequality $\beta>0$ follows from (\ref{Eq3}).
\bibitem{Notebt} It is interesting to note that the newly derived upper bound (\ref{Eq31}) on the mass
of the explosive scalar field is stronger than the bound (\ref{Eq5})
in the entire regime $a\lesssim 0.999481$.
\bibitem{Notesf} It is worth emphasizing again that
these stationary (marginally-stable) scalar field configurations are
characterized by the marginal (critical) frequency
$\omega_{\text{field}}=\omega_{\text{c}}\equiv m\Omega_{\text{H}}$
for superradiant scattering in the Kerr black-hole spacetime [see
Eq. (\ref{Eq4})].
\bibitem{Notenecs} That is, the inequality (\ref{Eq33}) provides a
necessary condition for the development of the superradiant
instabilities in the composed Kerr-black-hole-massive-scalar-field
system.
\bibitem{Notefs} It is worth noting that the expression on the r.h.s of (\ref{Eq33})
not only bounds the superradiant instability regime of the composed
Kerr-black-hole-massive-scalar-field system, but it also predicts
with a fairly good accuracy the exact (numerically computed
\cite{HerR}) scalar field masses which mark the onset of these
superradiant instabilities. For instance, from the data presented in
Table \ref{Table1} one learns that, for the $l=m=1$ mode, the
expression on the r.h.s of (\ref{Eq33}) agrees with the numerically
computed scalar field masses which mark the onset of the
superradiant instabilities to within $\sim 5\%$ in the regime
$a/M\lesssim0.9$. The agreement becomes even better for the higher
scalar modes. For instance, for the $l=m=10$ mode, the expression on
the r.h.s of (\ref{Eq33}) agrees with the numerically computed
scalar field masses which mark the onset of the superradiant
instabilities to within $\sim 2\%$ in the regime $a/M\lesssim0.95$.
\end{thebibliography}
\end{document}